\documentclass[review]{elsarticle}
\usepackage{lineno,hyperref}
\usepackage{color}
\usepackage[dvipsnames]{xcolor}
\usepackage[normalem]{ulem}
\pdfoutput=1
\modulolinenumbers[5]
\journal{Journal of \LaTeX\ Templates}

\begin{document}
\begin{frontmatter}
\title{$\eta$-pairing in the model with two-particle hybridization of conduction and localized  electrons}
\tnotetext[mytitlenote]{$\eta$-pairing in the model with two-particle hybridization of conduction and localized electrons}
\author{Igor N. Karnaukhov}
\address{G.V. Kurdyumov Institute for Metal Physics, 36 Vernadsky Boulevard, 03142 Kiev, Ukraine}
\fntext[myfootnote]{karnaui@yahoo.com}



\begin{abstract}
Within the framework of a model, that takes into account two-particle hybridization of conduction and localized electrons, the effective interaction between conduction electrons is calculated. It is shown that this interaction is attractive when the energy of the localized electron corresponding to the two-particle state lies in the conduction band above the Fermi energy. The magnitude of the attractive interaction is minimal for $\eta$-paired states of conduction electrons. We generalize the original $\eta$-pairing construction for the proposed model and show that the superconducting state can indeed be realized. 
\end{abstract}
 \begin{keyword}
\texttt  pairing \sep hybridization \sep high-temperature superconductivity
\end{keyword}
\end{frontmatter}
\section{Introduction}

Since the first discovery of high-temperature superconductivity \cite{0},
the question has arisen about the non-trivial nature of pairing  in the Fermi liquid. In contrast to traditional  singlet pairing of electrons with zero total moment (Cooper pairing), Yang \cite{Y} considered the pairing of conduction electrons with non-zero total moment, so-called $\eta$-pairing.
At $\eta$-pairing, electrons with momenta $\overrightarrow{k}$ and $\overrightarrow{\pi}- \overrightarrow{k}$ ($\overrightarrow{\pi}$ is vector) bind to each other, leading to the condensation of electron pairs with $\pi$-momentum, which in turn leads to spatially modulated superconductivity. Unfortunately $\eta$-pairing is realized in the Hubbard model with attractive interaction \cite{Y}. This rules out the use of this construction to explain the behaviour of real superconductors. Thus, the fundamental question about the nature of 
the pairing mechanism occurring in high-temperature superconductors remains
open.

In this article, we propose a fairly simple mechanism for the formation of effective attraction between conduction electrons. Based on the physical nature of this interaction, we have proposed a generalisation of the $\eta$-pairing construction. The author hopes that the proposed electron pairing mechanism can be implemented in real high-temperature superconductors.

 \section{The model Hamiltonian}

The model Hamiltonian  ${\cal H}={\cal H}_0+{\cal H}_v$ has the following form \cite{IK0}
\begin{eqnarray}
&&{\cal H}_0==- \sum_{\sigma}\sum_{j=1}^{L-1} (c^\dagger_{j\sigma}c_{j+1\sigma} +c^\dagger_{j+1\sigma}c_{j\sigma})+\varepsilon_g \sum_{\sigma}\sum_{j=1}^{L} n_{j\sigma}+ U\sum_{j=1}^{N} n_{j\uparrow}n_{j\downarrow},\nonumber \\&&
{\cal H}_{v}=v \sum_{j=1}^{L} (c^\dagger_{j\uparrow}c^\dagger_{j\downarrow}d_{j\uparrow}d_{j\downarrow}+
d^\dagger_{j\downarrow}d^\dagger_{j\uparrow}c_{j\downarrow}c_{j\uparrow}),
\label{eq:1}
\end{eqnarray}
where $c^\dagger_{j\sigma},c_{j\sigma}$ and  $d^\dagger_{j\sigma},d_{j\sigma}$ $(\sigma=\uparrow,\downarrow)$ are the Fermi operators of the conduction and localized electrons defined on the site $j$, $\varepsilon_g$ is the energy of  $d$-electron, $U$ is the  strength of the on-site Hubbard interaction for $d$-eelctrons, and $n_{j\sigma}=d^\dagger_{j\sigma}d_{j\sigma}$ and $m_{j\sigma}=c^\dagger_{j\sigma}c_{j\sigma}$ are the density operators for localized and conduction electrons. Parameter $v$  determines two-particle on-site hybridization  between localized and conduction electrons. Obviously,  the total  number of electrons $N_e=\sum_{\sigma}\sum_j (n_{j\sigma}+m_{j\sigma})$ and total spin $M=\frac{1}{2}\sum_j (n_{j\uparrow}+m_{j\uparrow}-n_{j\downarrow}-m_{j\downarrow})$ are conserved. The total number of lattice sites is $L$.

In the  single-impurity Anderson model, conduction electrons hybridize with localized electrons, which exist in three (one- and two-particle) states.  We will consider the case where the energy $\varepsilon_g$ corresponding to states of an electron with different spins is  outside the conduction band, so conduction electrons do not hybridize with $d$-electrons in these states (see Fig. 1, where the 1D model is shown for simplicity).  Conduction electrons hybridize with two localized electrons located at the same site, since the energy corresponding to this state lies within the conduction band. This interaction takes into account in the model Hamiltonian,, since the one-particle hybridization is not realised. 
The Hamiltonian (\ref{eq:1}) was proposed by the author in \cite{IK0}. 
In the model (\ref{eq:1}),  $d$-electrons are frozen, and  conduction electrons scatter by an effective potential $2 c(\textbf{k}_1,\textbf{k}_2)$, formed by the on-site interaction between conduction and localized electrons. The magnitude of the scattering potential is equal to $2 c(\textbf{k}_1,\textbf{k}_2)=\frac{v^2}{\epsilon(\textbf{k}_1) +\epsilon(\textbf{k}_2)-\epsilon_2}$, here $\epsilon(\textbf{k})$ the energy of a conduction electron with wave vector $\textbf{k}$ and $\epsilon_2= 2\varepsilon_g+U$ \cite{IK0}.  This work showed that when the energy of a localized electron $\epsilon_2/2$  corresponding to a two-particle state of $d$-electrons is far from the Fermi level, the 1D model can be solved using the Bethe ansatz. Without limiting ourselves to a chain, let us consider the behaviour of electron liquid in the system of arbitrary dimension, taking into account second-order perturbation theory and $\eta$-pairing construction.
\begin{figure}[tp]
      \centering{\leavevmode}
\begin{minipage}[h]{.5\linewidth}
\center{
\includegraphics[width=\linewidth]{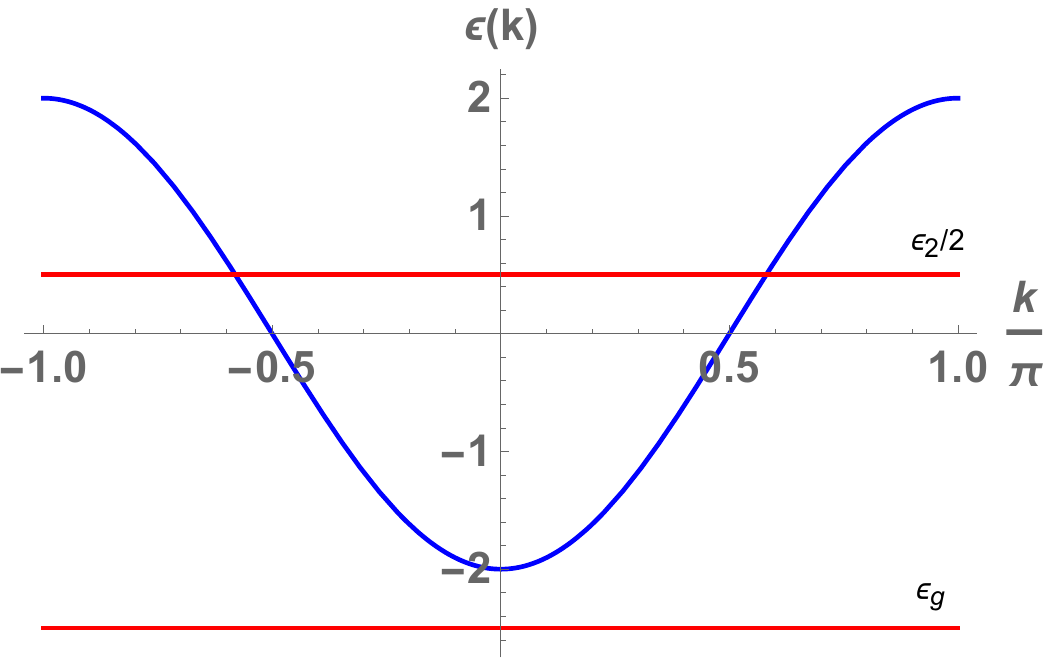} 
                  }
    \end{minipage}
\caption{The one-particle energies of conduction   $\epsilon (k)=-2\cos k$  (blue line) and localized electrons (red lines) as a function of the wave vector, $\epsilon_2=2\varepsilon_g +U$. 
} 
\label{fig:1}
\end{figure}
\begin{figure}[tp]
      \centering{\leavevmode}
\begin{minipage}[h]{.5\linewidth}
\center{
\includegraphics[width=\linewidth]{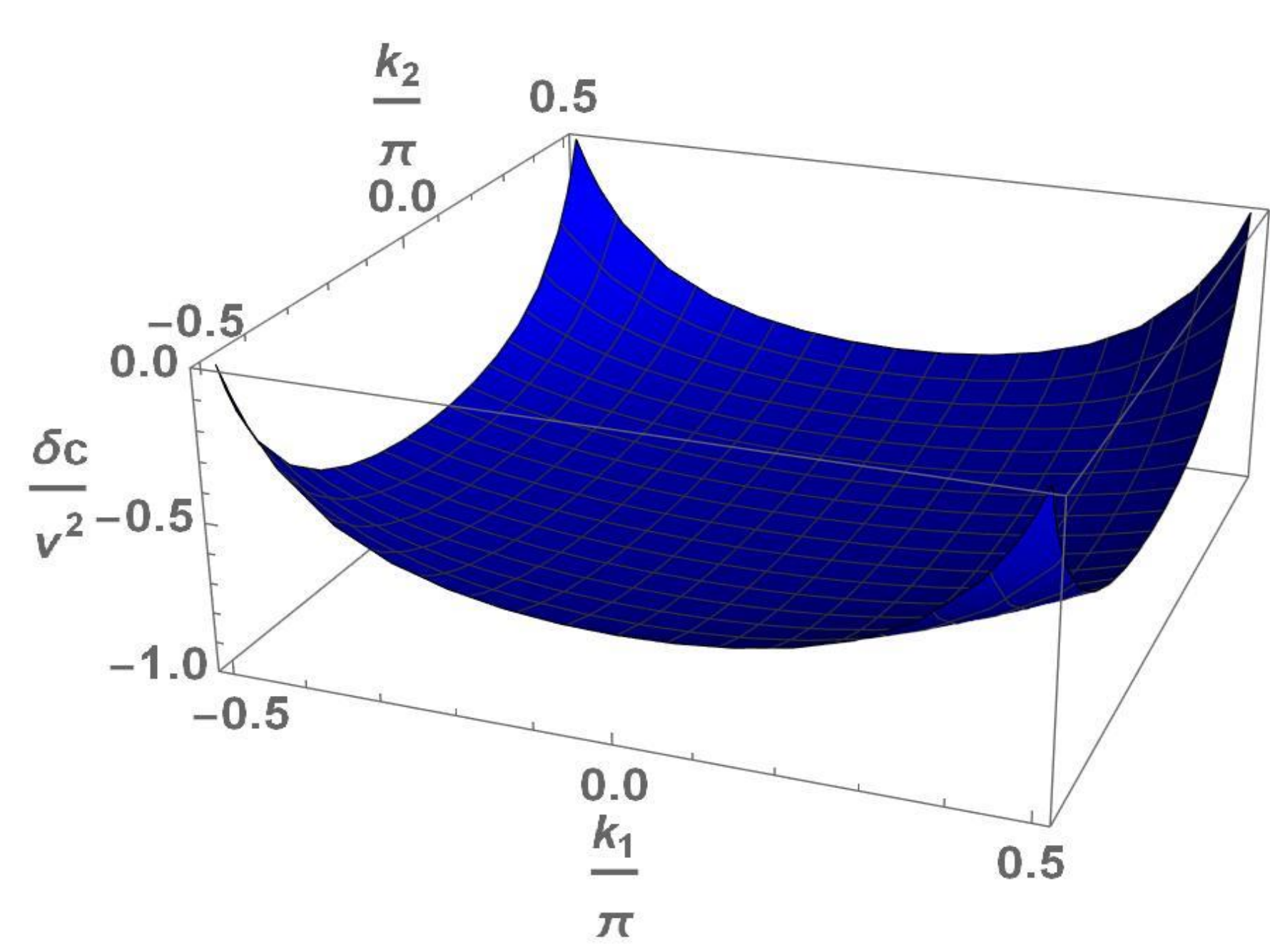} 
                  }
    \end{minipage}
\caption{$\delta c(k_1,k_2)/v^2=-\frac{1}{\epsilon_2}-\frac{1}{\epsilon (k_1)+\epsilon(k_2)-\epsilon_2}$ as a function of the wave vectors of scattered electrons $k_1$ and $k_2$ calculated for $\epsilon_2=1$, $\epsilon (k)=-2\cos k$. 
}
\label{fig:2}
\end{figure}

\section{Solution of the problem}


Let us redefine the model Hamiltonian (\ref{eq:1}) in the momentum presentation as ${\cal H}={\cal H}_{0}+{\cal H}_{v}$ 
\begin{eqnarray}
&&{\cal H}_0=\sum_{\sigma}\sum_\textbf{k} \epsilon (\textbf{k})c^\dagger_{\sigma}(\textbf{k})c_{\sigma}(\textbf{k})+\varepsilon_g \sum_{\sigma}\sum_\textbf{j} n_{\textbf{j}\sigma}+ U\sum_\textbf{j} n_{\textbf{j}\uparrow}n_{\textbf{j}\downarrow},\nonumber \\&&
{\cal H}_{v}=v \sum_\textbf{j } \sum_{\textbf{k}_1,\textbf{k}_2}[c^\dagger_{\uparrow}(\textbf{k}_1)c^\dagger_{\downarrow}(\textbf{k}_2) d_{\textbf{j}\uparrow}d_{\textbf{j}\downarrow} \exp(-i \textbf{k}_1 \textbf{j}-i \textbf{k}_2 \textbf{j})+\nonumber \\&&
d^\dagger_{\textbf{j}\downarrow}d^\dagger_{\textbf{j}\uparrow}c_{\downarrow}(\textbf{k}_2)c_{\uparrow}(\textbf{k}_1)\exp(i \textbf{k}_1 \textbf{j}+i \textbf{k}_2 \textbf{j})],
\label{eq:2}
\end{eqnarray}
where $c_{\textbf{j}\sigma}=\sum_{\textbf{k}}c_{\sigma}(\textbf{k})\exp(i \textbf{k j})$.

Using the Schrieffer-Wolff transformation, the Hamiltonian  (\ref{eq:2}) is defined in the second order interaction in the tradition form ${\cal H}_{eff}={\cal H}_0+\frac{1}{2}[S,{\cal H}_{v}]$, where $S$ is equal to 
\begin{eqnarray}
&&
S=\sum_{\textbf{k}_1,\textbf{k}_2}\frac {v}{\epsilon (\textbf{k}_1)+\epsilon (\textbf{k}_2)-\epsilon_2}\sum_{\textbf{j}}
[c^\dagger_{\uparrow}(\textbf{k}_1)c^\dagger_{\downarrow}(\textbf{k}_2)d_{\textbf{j}\uparrow}d_{\textbf{j}\downarrow} \exp(-i \textbf{k}_1 \textbf{j}-i \textbf{k}_2 \textbf{j})-
\nonumber\\&&
d^\dagger_{\emph{j}\downarrow}d^\dagger_{\textbf{j}\uparrow}c_{\downarrow}(\textbf{k}_2)c_{\uparrow}(\textbf{k}_1) \exp(i \textbf{k}_1 \textbf{j}+i \textbf{k}_2 \textbf{j})].
\label{eq:3}
\end{eqnarray}

The effective Hamiltonian ${\cal H}_{eff}={\cal H}_0+{\cal H}_{1}+{\cal H}_{2}$ includes three types of terms, here ${\cal H}_{1}$ and ${\cal H}_{2}$ are one-particle  and two-particle terms, according to the conduction electron operators. The term ${\cal H}_{1}$ renormalizes the one-particle energy of conduction electrons:
\begin{eqnarray}
&&
{\cal H}_{1}=\frac{v^2}{2}\sum_{\textbf{j}}\sum_{\textbf{k}_1,\textbf{k}_1',\textbf{k}_2,\textbf{k}_2'}
\frac{1}{\epsilon (\textbf{k}_1)+\epsilon (\textbf{k}_2)-\epsilon_2} 
\{n_{\textbf{j}\uparrow}n_{\textbf{j}\downarrow}[c^\dagger_{\uparrow}(\textbf{k}_1)c_{\uparrow}(\textbf{k}_1')\delta_{\textbf{k}_2,\textbf{k}_2'}-\nonumber\\&& 
c_{\downarrow}\textbf{(k}_2') c^\dagger_{\downarrow}(\textbf{k}_2)\delta_{\textbf{k}_1,\textbf{k}_1'}]
\exp(-i \textbf{k}_1 \textbf{j}+i \textbf{k}_1' \textbf{j}- i \textbf{k}_2 \textbf{j}+ i\textbf{k}_2' \textbf{j})+(1-n_{\textbf{j}\uparrow})(1-n_{\textbf{j}\downarrow})\nonumber\\&& 
[c^\dagger_{\uparrow}(\textbf{k}_1') c_{\uparrow}(\textbf{k}_1)\delta_{\textbf{k}_2,\textbf{k}_2'}-
c_{\downarrow}\textbf{(k}_2) c^\dagger_{\downarrow}(\textbf{k}_2')\delta_{\textbf{k}_1,\textbf{k}_1'}]
\exp(i \textbf{k}_1 \textbf{j}-i \textbf{k}_1' \textbf{j}+ i \textbf{k}_2 \textbf{j}- i\textbf{k}_2' \textbf{j})
\}.\nonumber\\
\label{eq:4}
\end{eqnarray}
The term ${\cal H}_{2}$ determines the scattering potential for conduction electrons, which is the result of the interaction between conduction and localized electrons:
\begin{eqnarray}
&&
{\cal H}_{2}=\frac{v^2}{2}\sum_{\textbf{j}}\sum_{\textbf{k}_1,\textbf{k}_1',\textbf{k}_2,\textbf{k}_2'}
\frac{1-n_{\textbf{j}\uparrow}-n_{\textbf{j}\downarrow}}{\epsilon (\textbf{k}_1)+\epsilon (\textbf{k}_2)-\epsilon_2} \nonumber\\&& 
[c^\dagger_{\uparrow}(\textbf{k}_1)c_{\uparrow}(\textbf{k}_1')c^\dagger_{\downarrow}(\textbf{k}_2)c_{\downarrow} (\textbf{k}_2')\exp(-i \textbf{k}_1 \textbf{j}+i \textbf{k}_1' \textbf{j}-i \textbf{k}_2 \textbf{j}+ i\textbf{k}_2' \textbf{j})
+\nonumber\\&& 
c^\dagger_{\uparrow}(\textbf{k}_1')c_{\uparrow}(\textbf{k}_1)c^\dagger_{\downarrow}\textbf{(k}_2') c_{\downarrow}(\textbf{k}_2)\exp(i \textbf{k}_1 \textbf{j}-i \textbf{k}_1' \textbf{j}+ i \textbf{k}_2 \textbf{j}- i\textbf{k}_2' \textbf{j})].
\label{eq:5}
\end{eqnarray}

In paper \cite{IK0} it has been shown, that the on-site two-particle interaction between conduction electrons does not depend on the density operators of localized electrons, because the localized electrons are frozen. As we noted above the one-particle states of localized electrons do not hybridize with conduction electrons, the one-particle states of $d$-electrons are forbidden. Thus $n_{\textbf{j}\sigma}=0$ in (\ref{eq:5}) and the effective on-site interaction between conduction electrons is equal to $2c(\textbf{k}_1,\textbf{k}_2)$. For energies of conduction electrons near the Fermi energy (when $\vert \varepsilon (\textbf{k})\vert <\frac{1}{2}\vert\epsilon_2 \vert $) the effective interaction between conduction electrons is repulsive for $\epsilon_2<0$ and attractive for $\epsilon_2>0$. 
For filling less than half, the magnitude of $c(\textbf{k}_1,\textbf{k}_2)$ has  minimal negative (for $\epsilon_2>0$) and maximal positive (for $\epsilon_2<0$) value for electrons with energies 
$\varepsilon (\textbf{k}_1)=\varepsilon (\textbf{k)}$ and $\varepsilon (\textbf{k}_2)=\varepsilon (\overrightarrow{\pi}-\textbf{ k}) =-\varepsilon(\textbf{k})$. To illustrate this, Fig. 2 shows the value $\frac{\delta c}{v^2}=2c(\textbf{k},\overrightarrow{\pi}-\textbf{ k})/v^2-2c(\textbf{k}_1,\textbf{k}_2)/v^2=-\frac{1}{\epsilon_2}-\frac{1}{\epsilon (k_1)+\epsilon(k_2)-\epsilon_2}$ as a function of wave vectors of electrons, calculated at $\epsilon_2=1$ in the 1D version of the model. For $\epsilon_2>0$, electrons with wave vectors $\textbf{k}$ and $\overrightarrow{\pi}-\textbf{ k}$ form a bound state with a binding energy equal to $-\frac{v^2}{\epsilon_2}$, thus referring to $\eta$-pairing \cite{Y}. The binding energy increases as the energy of level $\frac{\epsilon_2}{2}$ approaches the Fermi energy.

When the energy of the two-particle $d$-level lies above the Fermi energy  ($d$-state is empty $n_{\textbf{j}\uparrow} n_{\textbf{j}\downarrow}=0$ at $v=0$), conduction electrons  hybridize with $d$-electrons, which leads to a decrease in their effective charge. As a result, the effective interaction between conduction electrons is attractive. For $\varepsilon_2 <0$, the  two-particle $d$ level is filled ($d$-state is filled  $n_{\textbf{j}\uparrow} n_{\textbf{j}\downarrow}=1$ at $v=0$), in which case the effective charge of conduction electrons increases. This, in turn, leads to repulsion of conduction electrons. 

\section{$\eta$-pairing construction}

In the model (\ref{eq:1}) $\eta$-pairing (Yangs definition \cite{Y}) is determined by two operators $\eta=\sum_\textbf{j} \exp (i \overrightarrow{\pi} \overrightarrow{j})c^\dagger_{\textbf{j}\uparrow}c^\dagger_{\textbf{j}\downarrow}$  and 
$\mu=-\sum_\textbf{j} \exp (i \overrightarrow{\pi} \overrightarrow{j})d^\dagger_{\textbf{j}\uparrow} d^\dagger_{\textbf{j}\downarrow}$. These operators describe pairs of conduction and localized electrons, as these states are bounded  together through interaction. The operators satisfy the relations
\begin{eqnarray}
&&[{\cal H},\eta]=v\mu-v\sum_\textbf{j} \exp (i \overrightarrow{\pi} \overrightarrow{j}) d^\dagger _{\textbf{j}\downarrow} d^\dagger_{\textbf{j}\uparrow}( m_{\textbf{j}\uparrow} +m_{\textbf{j}\downarrow}),\nonumber \\ 
&&[{\cal H},\mu]=\epsilon_2\mu+v\eta-v \sum_\textbf{j} \exp (i \overrightarrow{\pi} \overrightarrow{j}) c^\dagger_{\textbf{j}\uparrow}c^\dagger_{\textbf{j}\downarrow}(n_{\textbf{j}\uparrow}+ n_{\textbf{j}\downarrow}).
\label{eq:6}
\end{eqnarray}
According to (\ref{eq:1}), $d$-electrons are frozen, and in the case of $\eta$-pairing, the bound states of conduction electrons (electron pairs) are also frozen. We can define the condition for particle conservation at the lattice cell as $2m_{\textbf{j}\uparrow}m_{\textbf{j}\downarrow}+2n_{\textbf{j}\uparrow}n_{\textbf{j}\downarrow}=\rho$, where $\rho$ is the local electron density.
As we noted above $d$-electrons in one-particle states do not interact with conduction electrons, thus $n_{\textbf{j}\uparrow}=n_{\textbf{j}\downarrow}=0$ in the second Eq of (\ref{eq:6}). The similar relations are valid for frozen conduction electrons $m_{\textbf{j}\uparrow}=m_{\textbf{j}\downarrow}=0$ in the first Eq of (\ref{eq:6}). Taking the above into account, the system of Eqs (\ref{eq:6}) is simplified
\begin{eqnarray}
&&[{\cal H},\eta]=v\mu,\nonumber \\ 
&&[{\cal H},\mu]=\epsilon_2\mu+v\eta .
\label{eq:7}
\end{eqnarray}
and the energy of electron pair is equal to ${\cal E}=\frac{1}{2}(\epsilon_2 \pm \sqrt{\epsilon_2^2+ 4 v^2})$. In the case $\epsilon_2 >>v>0$ the energy is negative and equal to ${\cal E}=-\frac{v^2}{\epsilon_2}$, for another limit $0<\epsilon_2 <<v$ ${\cal E}=-v$.
The one-pair eigenstate of the Hamilton is $\vert \Psi_1=\eta \vert vac>$, ${\cal E}$, the energy $N{\cal E}$ corresponds to $N$-pairs state $\vert \Psi_N=\eta^N \vert vac>$.

The off-diagonal long-range order of this quantum state is calculated in \cite{Y} as
\begin{eqnarray}
&&<\Psi_N\vert \eta_{j_1} \eta^\dagger_{j_2}\vert \Psi_N>\sim\frac{N}{L}\left(1-\frac{N}{L}\right)
\end{eqnarray}
where  $\vert j_1-j_2\vert \to \infty$. In the thermodynamic limit, $N,L \to \infty$, where $N/L = n$, the off-diagonal long-range order is equal to $n(1-n)$, and for the case of $\eta$-pairing with half-filling $n = 1/2$, it reaches its maximum.

\section{Conclusion}
Taking into account second-order perturbation theory, we calculated the effective Hamiltonian, which determines the effective two-particle interaction between conduction electrons through their hybridization with two-particle states of localized electrons.
It has been shown that attractive interactions between conduction electrons occur when conduction electrons hybridize with localized electrons in a two-particle state,  the energy of localized electron corresponding to this state lies above the Fermi energy. The maximum attraction between conduction electrons occurs when they form an $\eta$-pair state.
The proposed mechanism for forming effective interaction between conduction electrons through their two-particle hybridization with localized electrons allows the implementation of the $\eta$-pairing construction to explain   high-temperature superconductivity in real systems.

\subsection*{Author contributions statement}
I.K. is the author of the manuscript
\subsection*{Additional information}
The author declares no competing financial interests. 
\subsection*{Availability of Data and Materials}
All data generated or analysed during this study are included in this published article.

\end{document}